\documentclass[fleqn,usenatbib]{mnras}

\usepackage{newtxtext,newtxmath}

\usepackage[T1]{fontenc}

\DeclareRobustCommand{\VAN}[3]{#2}
\let\VANthebibliography\thebibliography
\def\thebibliography{\DeclareRobustCommand{\VAN}[3]{##3}\VANthebibliography}

\usepackage{graphicx}	
\usepackage{amsmath}	

\title[On the high linearly polarized pulsar signals]{The Thousand-Pulsar-Array programme on MeerKAT XIV: On the high linearly polarized pulsar signals}

\author[Johnston et al.]{
Simon Johnston$^{1}$\thanks{Email:Simon.Johnston@csiro.au},
Dipanjan Mitra$^{2,3}$,
Michael~J. Keith$^{4}$,
Lucy~S.~Oswald$^{5,6}$,
Aris Karastergiou$^{5}$\\
$^{1}$Australia Telescope National Facility, CSIRO Space and Astronomy, PO Box 76, Epping NSW 1710, Australia\\
$2$National Centre for Radio Astrophysics, Tata Institute for Fundamental Research, Post Bag 3, Ganeshkhind, Pune 411007, India \\
$3$Janusz Gil Institute of Astronomy, University of Zielona Góra, ul. Szafrana 2, 65-516 Zielona Góra, Poland\\
$^4$Jodrell Bank Centre for Astrophysics, Department of Physics and Astronomy, University of Manchester, Manchester M13 9PL, UK\\
$^{5}$Department of Astrophysics, University of Oxford, Denys Wilkinson Building, Keble Road, Oxford OX1 3RH, UK\\
$^{6}$Magdalen College, University of Oxford, Oxford OX1 4AU, UK\\
}
\date{Accepted XXX. Received YYY; in original form ZZZ}

\pubyear{2024}

\begin{document}
\label{firstpage}
\pagerange{\pageref{firstpage}--\pageref{lastpage}}
  \maketitle

\begin{abstract}
The S-shaped swing of the linear polarization position angle (PPA) observed in many pulsars can be interpreted by the rotating vector model (RVM). However, efforts to fit the RVM for a large sample of pulsars observed with the MeerKAT telescope as a part of the Thousand-Pulsar-Array (TPA) programme, only succeeded for about half the cases. High time-resolution studies suggest that the failed cases arise due to the presence of orthogonal polarization modes, or highly disordered distribution of PPA points. One such example is PSR~J1645--0317. Recently it has been shown that the RVM can be recovered in this pulsar by using only time samples which are greater than 80\% linearly polarized. In this work we test this novel approach on the brightest 249 pulsars from the TPA sample, of which 177 yield sufficient highly polarized samples to be amenable to our method. Remarkably, only 9 of these pulsars (5\%) now fail to fit the RVM as opposed to 59\% from the original analysis. This result favours the paradigm that the underlying mechanism is coherent curvature radiation.
\end{abstract}




\section{Introduction}
\label{intro}
\cite{1969ApL.....3..225R} first observed the characteristic S-shaped swing of the linear polarization position angle (PPA) across the pulse profile in the Vela pulsar and proposed the rotating vector model (RVM) to explain the observations. In the RVM the PPA swing results due to an emission mechanism where the linearly polarized emission traces the variation in the dipolar magnetic field line planes. As the star rotates, the observer cuts different dipolar magnetic field lines, which then produces the characteristic S-shaped swing. \cite{1969PASA....1..254R} suggested that the underlying emission mechanism was vacuum curvature radiation, with emission originating close to the magnetic pole of the neutron star and the electric field vectors lying in the dipolar magnetic field line plane. In subsequent studies however it was found that the electric field emerging from the Vela pulsar lies perpendicular to the dipolar magnetic field planes \citep{2001ApJ...556..380H,2001ApJ...549.1111L}, and hence the vacuum curvature radiation model for the Vela pulsar had to be abandoned. 

In due course as more pulsars were discovered, polarization studies have been reported for a large sample of pulsars (see e.g. \citealt{1971ApJS...23..283M,1998MNRAS.301..235G,1999ApJS..121..171W,2008MNRAS.388..261J,2009ApJS..181..557H,2018MNRAS.474.4629J,2023MNRAS.524.5042R,2023RAA....23j4002W,2023MNRAS.520.4582P}). The majority of these studies are so-called average profile studies, where the Stokes parameters are averaged for a few hundred to a few thousand pulses. These studies often reveal that the smooth PPA traverse is interrupted by jumps of 90\degr\ due to the presence of orthogonal polarization modes (OPM; \citealt{1975PASA....2..334M, 1976Natur.263..202B}). The PPA behaviour of the larger sample of pulsars showed that the RVM is a reasonable model for the normal pulsar population i.e. pulsars with periods longer than $\sim$ 0.05 second, and in this work we focus on this population. While the RVM is a good fit to many normal pulsars, there are equally a large number of pulsars where the RVM fails. The most comprehensive study in this regard is by \cite{2023MNRAS.520.4801J} (hereafter J23), where they tested the validity of the RVM against the average PPA behaviour for 850 pulsars, and found that the RVM could only be fitted to just over half of the sample, even when the orthogonal jumps are accounted for.
\begin{figure*}
\begin{center}
\begin{tabular}{cc}
\includegraphics[width=8cm]{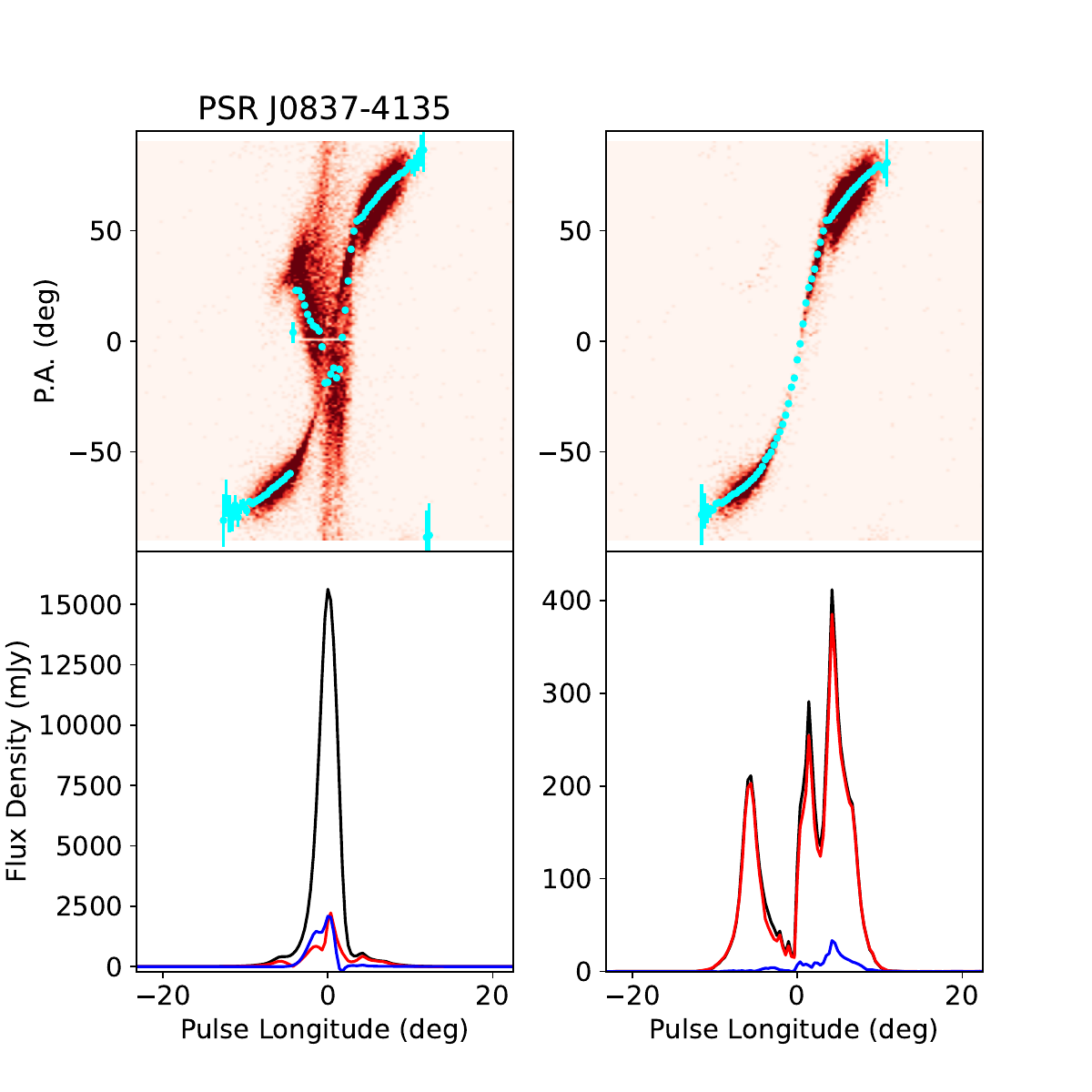}  &
\includegraphics[width=8cm]{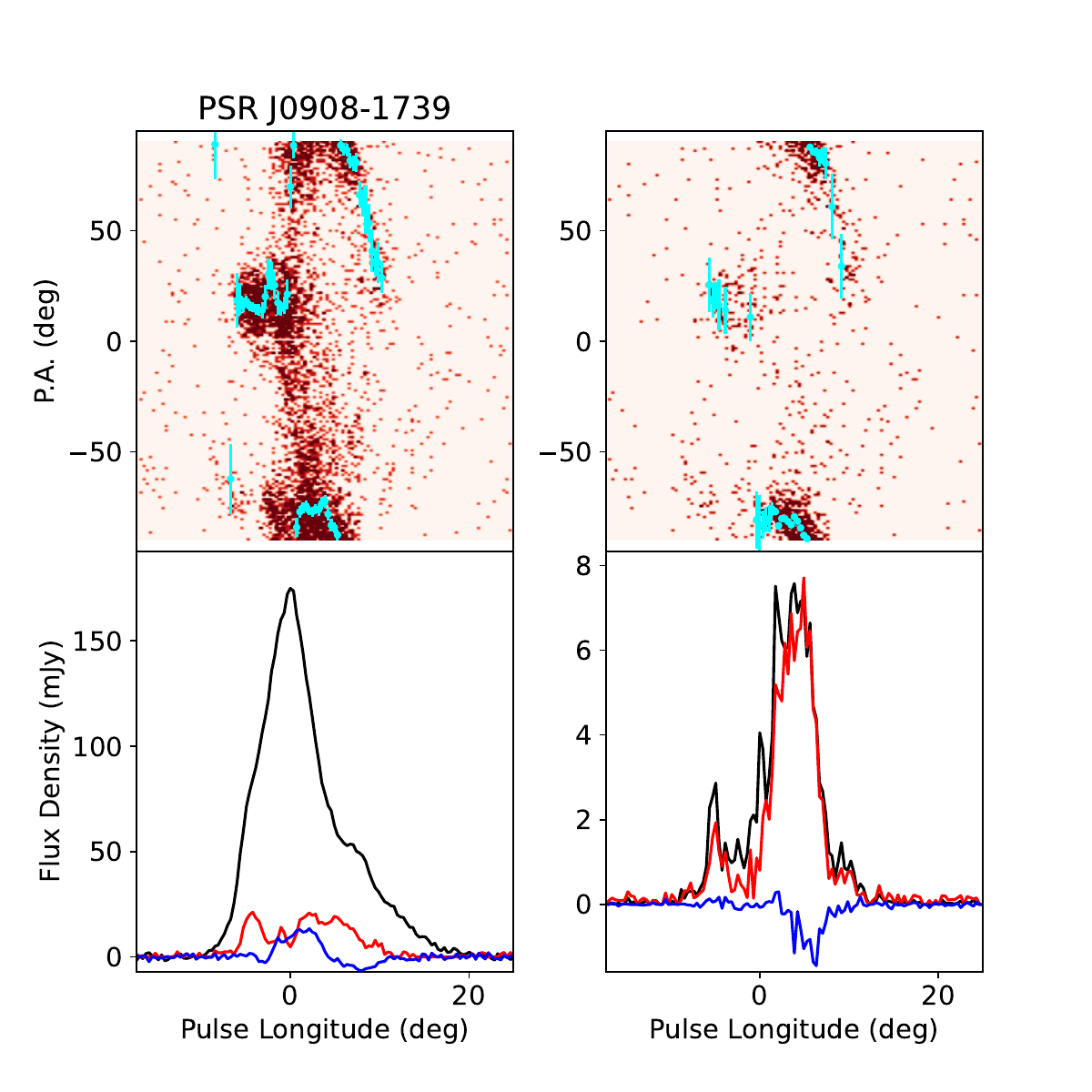}  \\
\includegraphics[width=8cm]{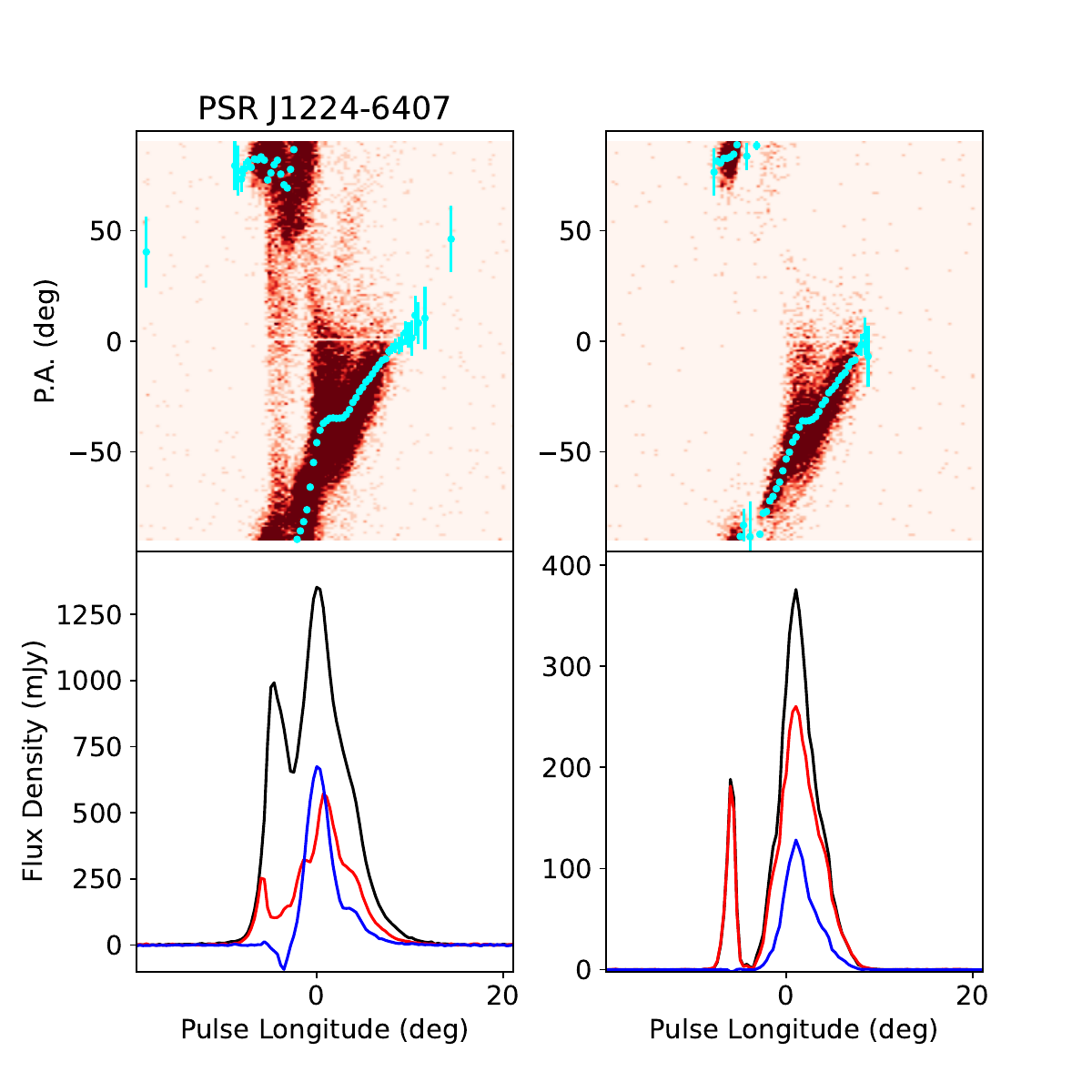}  &
\includegraphics[width=8cm]{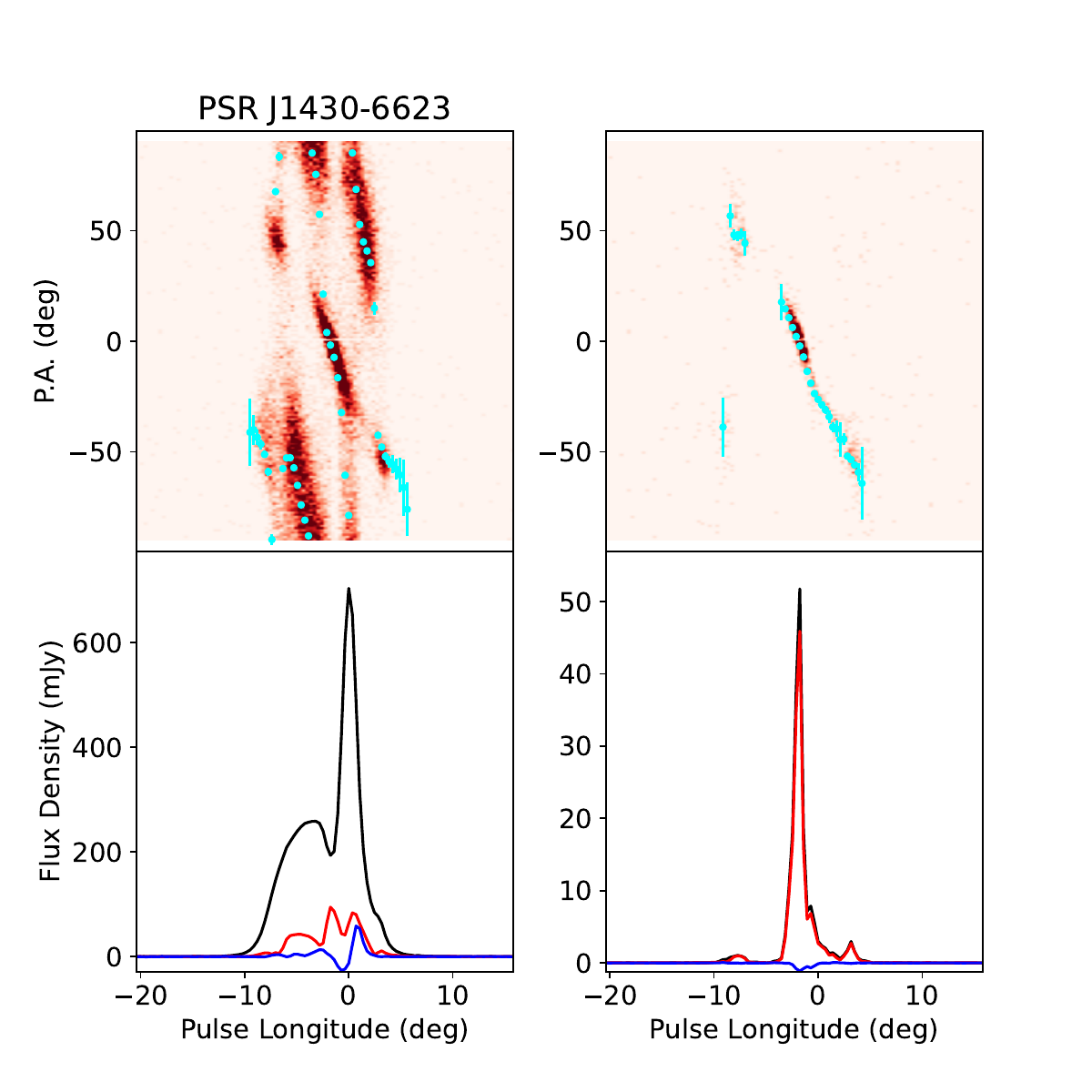}  \\
\end{tabular}
\end{center}
\caption{Results for four pulsars with non-RVM PPA tracks after integrating all the single pulses. Each pulsar's data is represented by four panels. The top left panel shows in cyan the position angle of the linear polarization in the integrated profile superposed on a heat-map of all the individual position angle points. The bottom left panel show the integrated profile (black), the linear polarization (red) and the circular polarization (blue). The right hand panels depict the same scheme but only for those samples which are more than 80\% polarized.}
\label{fig1}
\end{figure*}

Instead of integrating over many rotations of the pulsar, examination of the polarization on a continuous train of time-samples (typically of duration 1~ms or less) often provides a better understanding of the PPA behaviour. \citet{1995MNRAS.276L..55G} showed that in a pulsar with a complex non-RVM like PPA traverse, the time-sampled PPA distribution clearly showed two parallel orthogonal RVM tracks.  However the many PPA values which did not conform to the RVM were then speculated to be due to mode mixing and propagation effects. This aspect of OPM averaging has also been seen in other studies of bright pulsars (e.g. \citealt{1995JApA...16..327R,2004ApJ...606.1167R,2016MNRAS.460.3063M}). A handful of these time-sampled polarization emission surveys have been carried out for normal pulsars (\citealt{1975ApJ...196...83M,1984ApJS...55..247S, 2011ApJ...727...92M,2016ApJ...833...28M, 2019MNRAS.489.1543O}), where often RVM-like OPM tracks are seen in many pulsars across different frequencies.
These OPM tracks are usually thought to be associated with the natural modes i.e. the extraordinary (X) and ordinary (O) propagation modes of pair plasma embedded in the strong magnetic field \citep{1976ApJ...204L..13C,1977PASA....3..120M} .

There exists another category of pulsars which are significant in number, where the time-sampled PPA appear to show no organized pattern and the corresponding average PPA has a complex non-RVM like behaviour. Recently \cite{2023MNRAS.tmpL..22M} (MMB23 hereafter) claimed that the high linearly polarized time samples must correspond to purely linearly polarized X and O mode of the strongly magnetized pair plasma, and if the underlying emission mechanism is excited due to coherent curvature radiation (CCR hereafter), then these time samples across the pulse should reproduce the RVM. This line of reasoning is similar to an earlier study by \cite{2009ApJ...696L.141M} and \cite{2014ApJ...794..105M} where they demonstrated that the PPA corresponding to the high linearly polarized subpulses in several pulsars follow the average RVM like PPA traverse. MMB23 applied the high linearly polarized time sample criterion to PSR~J1645$-$0317, which has a disordered single pulse PPA distribution and complex non-RVM average PPA traverse. Remarkably they found that the high linearly polarized samples followed the PPA track which can be fitted with the RVM accurately. In this case only one PPA track was clearly visible, and a hint of presence of high linearly polarized orthogonal PPAs for a small range of pulse longitude was seen. MMB23 surmised that the low linearly polarized samples dominate the disordered non-RVM PPAs, presumably due to mixing of the modes and propagation effects.
 
The high linearly polarized sample criteria can hence be employed to a larger dataset to check two predictions of MMB23: firstly in pulsars with disordered PPA the high linearly polarized samples should follow the RVM and secondly there must be several cases where these samples follow the two tracks of X and O mode. The sample set used by J23 is particularly suitable for this exercise, since for many pulsars high quality, high time-resolution data have also been recorded. In this paper we report an analysis of high linearly polarized samples from the TPA observations. In section~\ref{sec2} we define the sample selection, section~\ref{sec3} we describe the analysis technique and present the results and in section~\ref{sec4} we state our conclusions.

\section{Sample selection}
\label{sec2}
The Thousand Pulsar Array Program (TPA) on MeerKAT has observed in excess of 1200 pulsars \citep{2020MNRAS.493.3608J}. The integration time on each of the pulsars was generally sufficient to produce at least 1000 single pulses \citep{2021MNRAS.505.4456S}. Details of the TPA observing, processing, calibration and analysis can be found in \cite{2021MNRAS.505.4483S}, \cite{2021MNRAS.508.4249P} and \cite{2023MNRAS.520.4562S}. In brief the observing band runs from 896 to 1671~MHz subdivided into 1~MHz channels. Typically 1024 phase bins per pulsar period are formed and the data are both flux and polarization calibrated.

In this paper we used high time-resolution data for 1202 pulsars. From the J23 paper, these pulars are classified as follows: 411 follow the RVM, 342 are non-RVM, 70 are `flat' i.e no significant PPA changes across the pulse profile but still can be fitted with RVM, 20 are interpulses and for 359 no RVM fit was attempted. In J23, the signal-to-noise ratio (s/n) of the linear polarization in the integrated profile was critical, here however, we require good s/n for each single pulse and we need sufficient single pulses to build up the statistics. We therefore restricted the sample to (a) pulsars with more than 500 pulses and (b) at least 60\% of the single pulses had to have s/n more than 5-$\sigma$ in Stokes $I$. This resulted in a sample of 249 pulsars.
\begin{figure*}
\begin{center}
\begin{tabular}{cc}
\includegraphics[width=8cm]{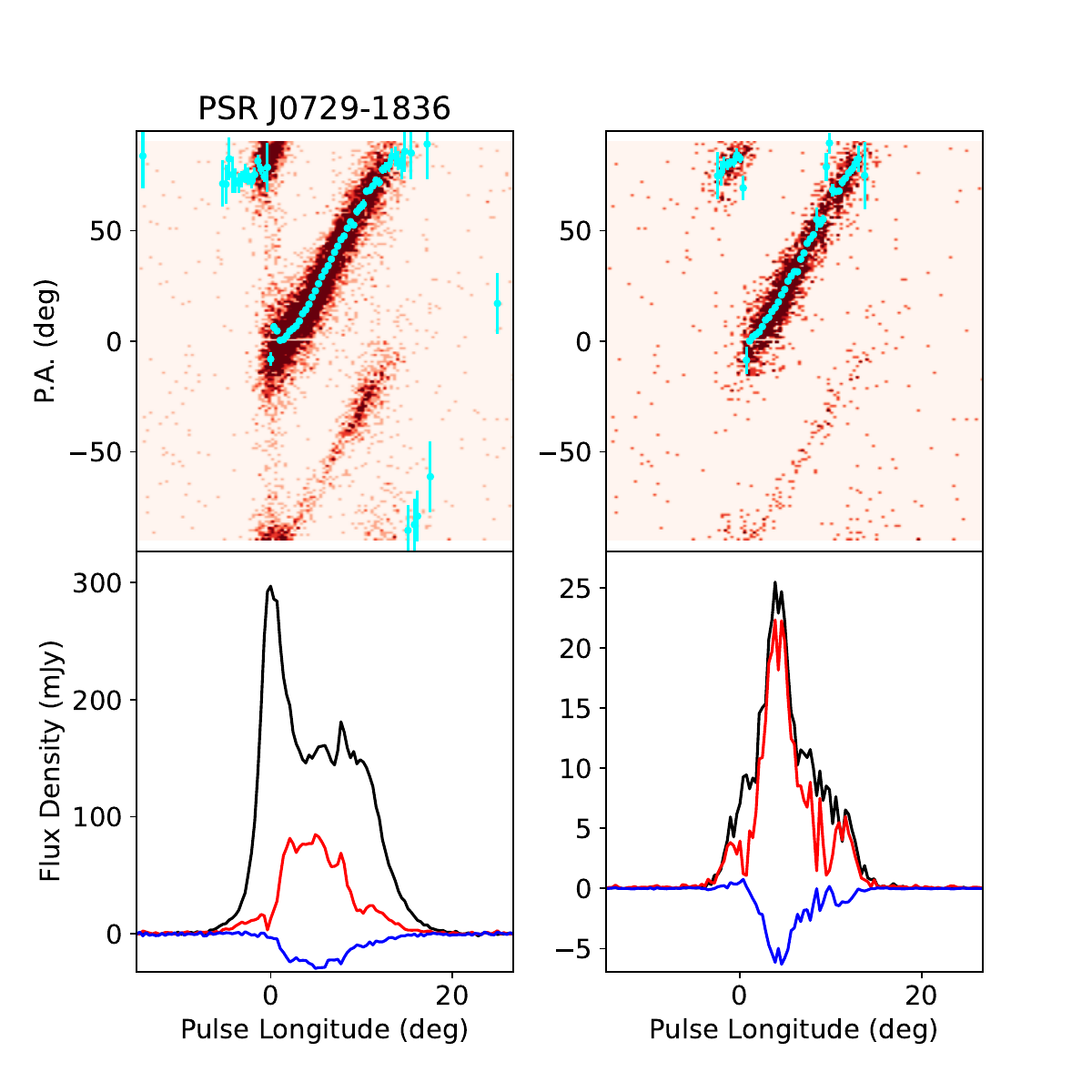}  &
\includegraphics[width=8cm]{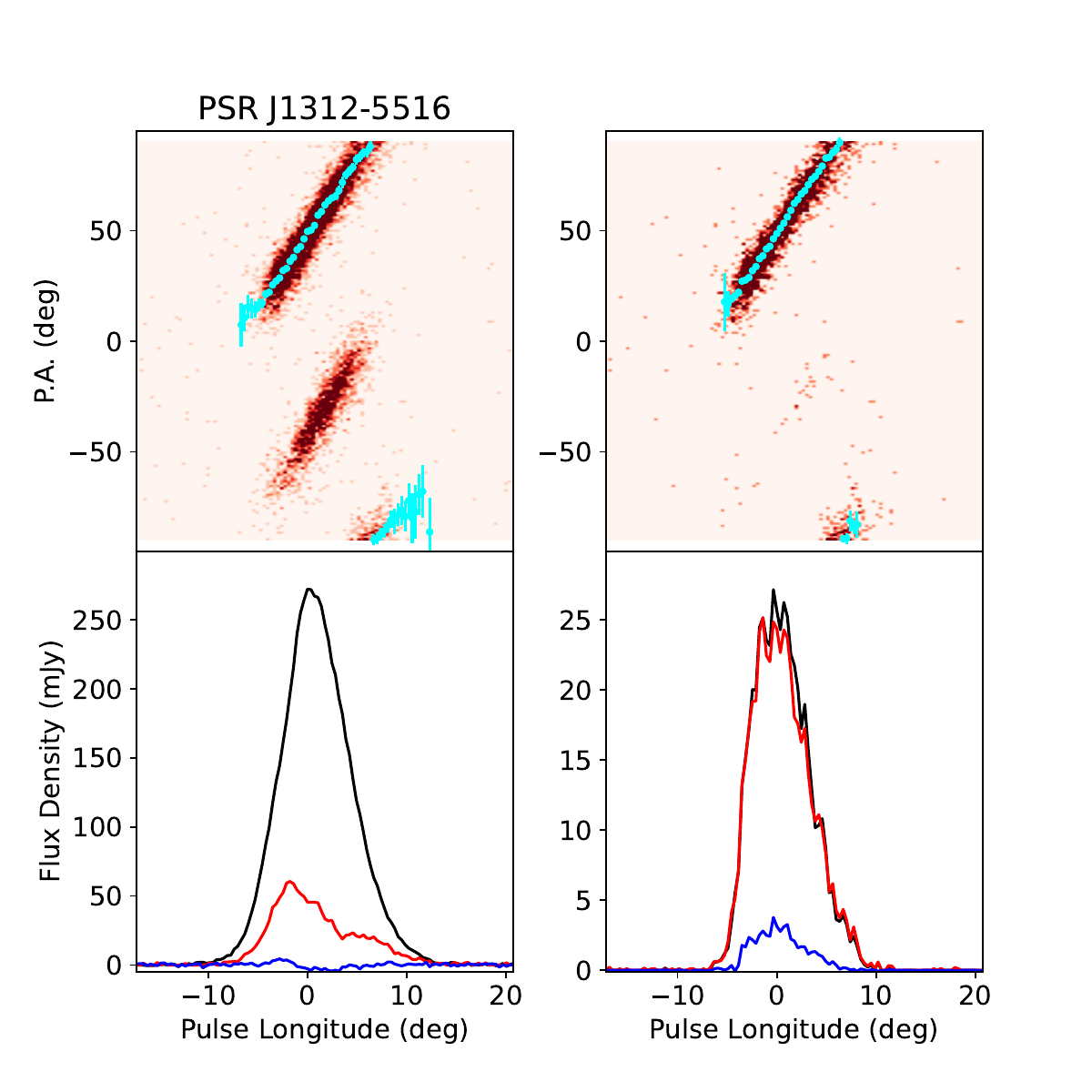}  \\
\includegraphics[width=8cm]{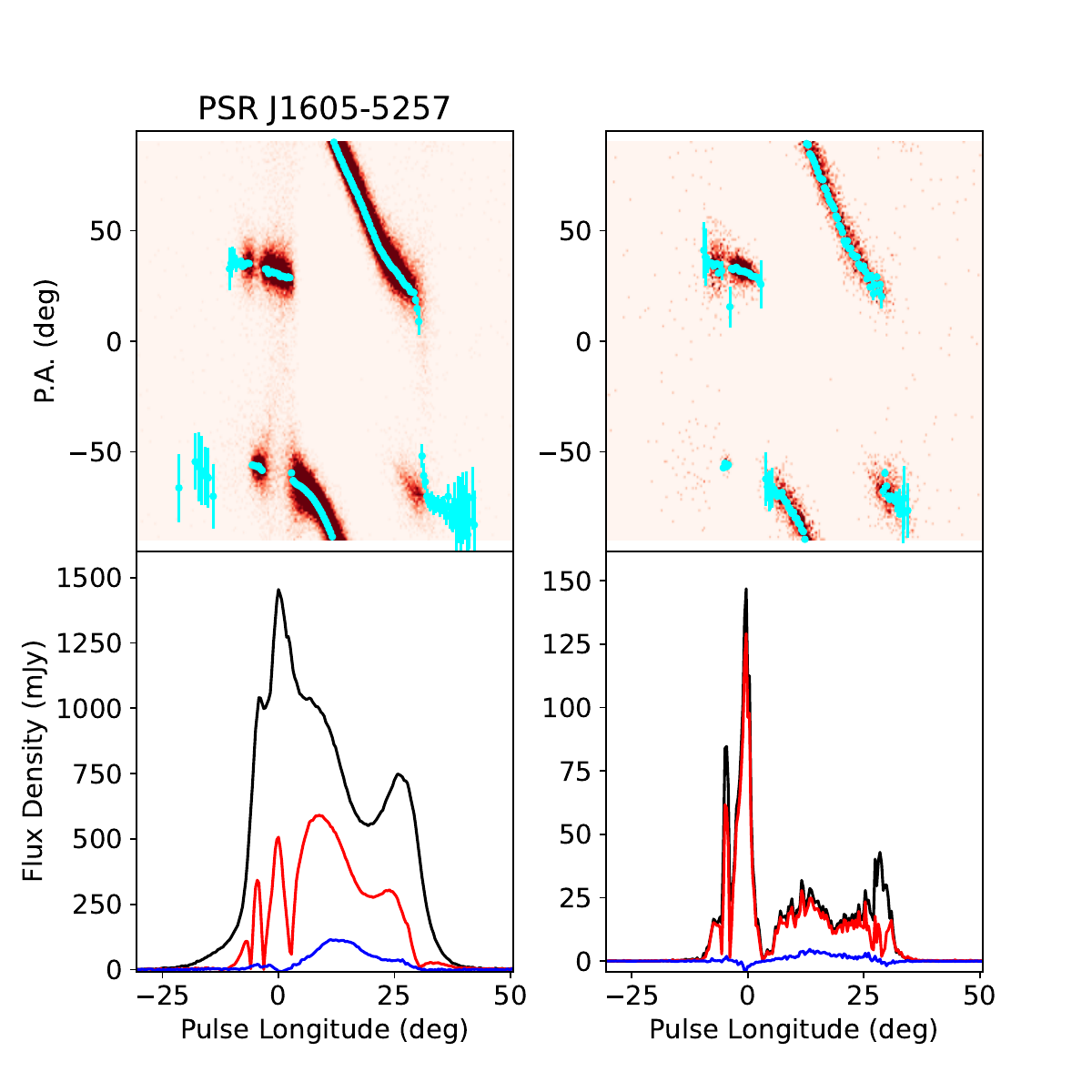}  &
\includegraphics[width=8cm]{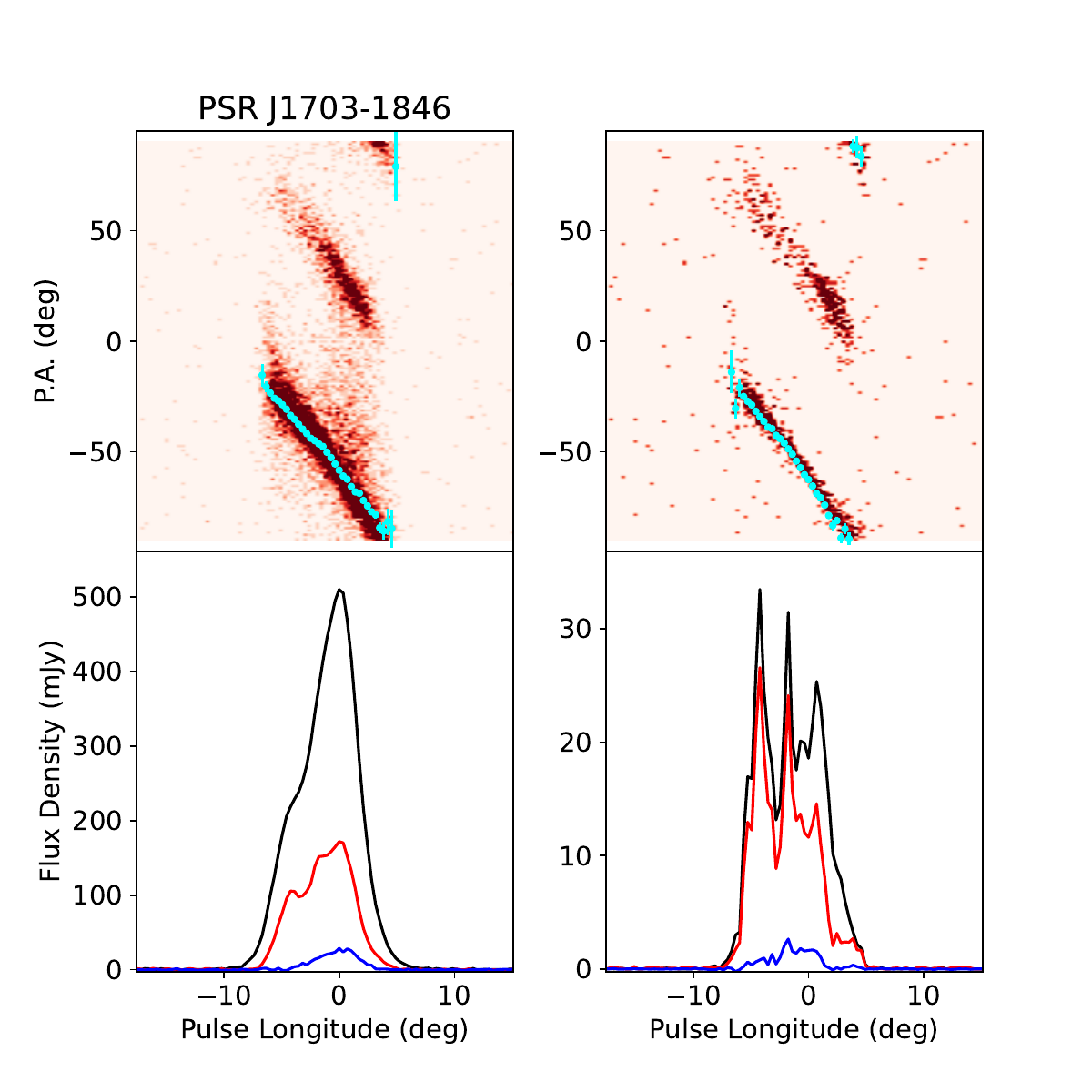}  \\
\end{tabular}
\end{center}
\caption{Results for four pulsars showing RVM tracks with orthogonal jumps. Each pulsar's data is represented by four panels. The top left panel shows in cyan the position angle of the linear polarization in the integrated profile superposed on a heat-map of all the individual position angle points. The bottom left panel show the integrated profile (black), the linear polarization (red) and the circular polarization (blue). The right hand panels depict the same scheme but only for those samples which are more than 80\% polarized.}
\label{fig2}
\end{figure*}

\section{Results and Analysis}
\label{sec3}
\subsection{Forming the highly polarized profiles}
The analysis proceeded as follows. First we summed together all the data modulo the pulsar spin period to produce the average pulse profile and set the on-pulse window. At the same time, for each phase bin within the pulse window we computed the PPA provided the linear polarization was more than 3.5-$\sigma$ above the noise. Then we made a second pass through the high time-resolution data, this time only recording samples which are greater than 80\% linearly polarized. We kept track of the number of occurrences of such samples as a function of pulse longitude. As before, we computed the PPA for each of these samples. Finally, we produce an integrated profile from these high polarized samples and an integrated PPA swing. We stress that we deal with individual time samples, we do not consider single pulses as a whole nor any specific sub-structure within the single pulse profile.

In Figure~\ref{fig1} we show the output from four pulsars, all of which were classified as non-RVM in J23. The most noticeable aspect here is that on application of the high polarized time sample criterion, the disordered PPA distribution (seen in the left top heat map) vanishes, and a more ordered narrowly confined PPA distribution is observed (seen in the right top heat map). Consequently the distortions in the average PPA tracks (cyan points in the left top panel) smooth out, and more RVM-like tracks appear (cyan points in the right top panel). The effect is striking in PSR~J0837$-$4135, where the classical RVM-like PPA becomes apparent. In PSR~J0908--1739, the OPM in the leading part of the profile gets clearly resolved. In PSR~J1224--6407, while most of the disordered PPA distribution vanishes leaving behind a narrowly defined PPA track, one can still see some excess spread of PPA towards the center of the profile which results in a small kinky feature in average PPA. In PSR~J1430--6623 the several PPA patches almost disappear in the high polarized case and only a single RVM like PPA track is seen. 

In Figure~\ref{fig2} we show four examples of pulsars that were classified as RVM in J23. After applying the high polarized criteria the average PPA traverse for these pulsars still follows the RVM. The high polarized PPA distributions show assorted properties: in PSRs~J0729--1836 and J1703--1846 the two parallel OPM tracks are seen in both the full and high polarized case; in PSR~J1312--5516 two tracks are seen in the full data while only one track is visible in high polarized case; in PSR~J1605--5257 the multiple OPM jumps in the full data are reproduced in the high polarized data.

It should be noted that, for low $\dot{E}$ pulsars, only a few percent of the samples go into making the profiles shown in Figures~\ref{fig1} and \ref{fig2}. This would imply that part of the depolarization is due to the spread of the PPA, but most of the depolarization must already occur internally in the magnetosphere, before the emission detaches.

\begin{table}
\caption{Results for the 249 pulsars analysed here. The first column is the classifcation given by the RVM fitter (see J23 for details). The second and third column denote how many pulsars fall into each class according to J23 and from the present work.}
\label{tab:results}
\begin{tabular}{lrr}
\hline
Class & J23 & current work\\
\hline
RVM & 90 & 146\\
Flat & 11 & 9 \\
non-RVM & 146 & 22 \\
No Pol & 2 & 72 \\
\hline
\end{tabular} 
\end{table}
\subsection{RVM fitting}
\label{secrvm}
The same RVM fitter that was used in J23 was now used to fit the high linearly polarized average PPA tracks for all 249 pulsars with the results given in Table~\ref{tab:results}. None of the pulsars previously classified as RVM became non-RVM after selecting the high polarized samples. In 72 of the 249 pulsars, there were not sufficient PPA points for the fitter to proceed (i.e. there were not sufficient highly polarized samples). Of the 177 remaining pulsars we found that only 22, i.e. 12\%, failed to give good fits with $\chi^2$ values less than 3 and hence were classified as non-RVM. This is a significant improvement over the results of J23, where RVM failed to fit to the average PPA for 59\% of the pulsars under consideration here.

We carefully scrutinized the 22 cases which the fitter classified as `non-RVM'. These cases may be influenced by our analysis method, where due to practical considerations for retaining sufficient time samples we have defined the high linearly polarized sample to be above 80\%, whereas MMB23 suggest that the RVM will be reproduced only by 100\% linearly polarized samples. The lower fractional polarization can be a result of mixing of the X and O modes, and it is possible that this leads to deviation from the RVM. We tested this by re-analysing the data for these 22 pulsars when setting the threshold to 90\% linear rather than 80\%. This indeed reduces the number of non-RVM cases to 9. We surmise that more sensitive observations and a larger number of pulsar rotations would be beneficial to sort out the remaining cases.

\begin{figure}
\begin{center}
\begin{tabular}{c}
\includegraphics[width=8cm]{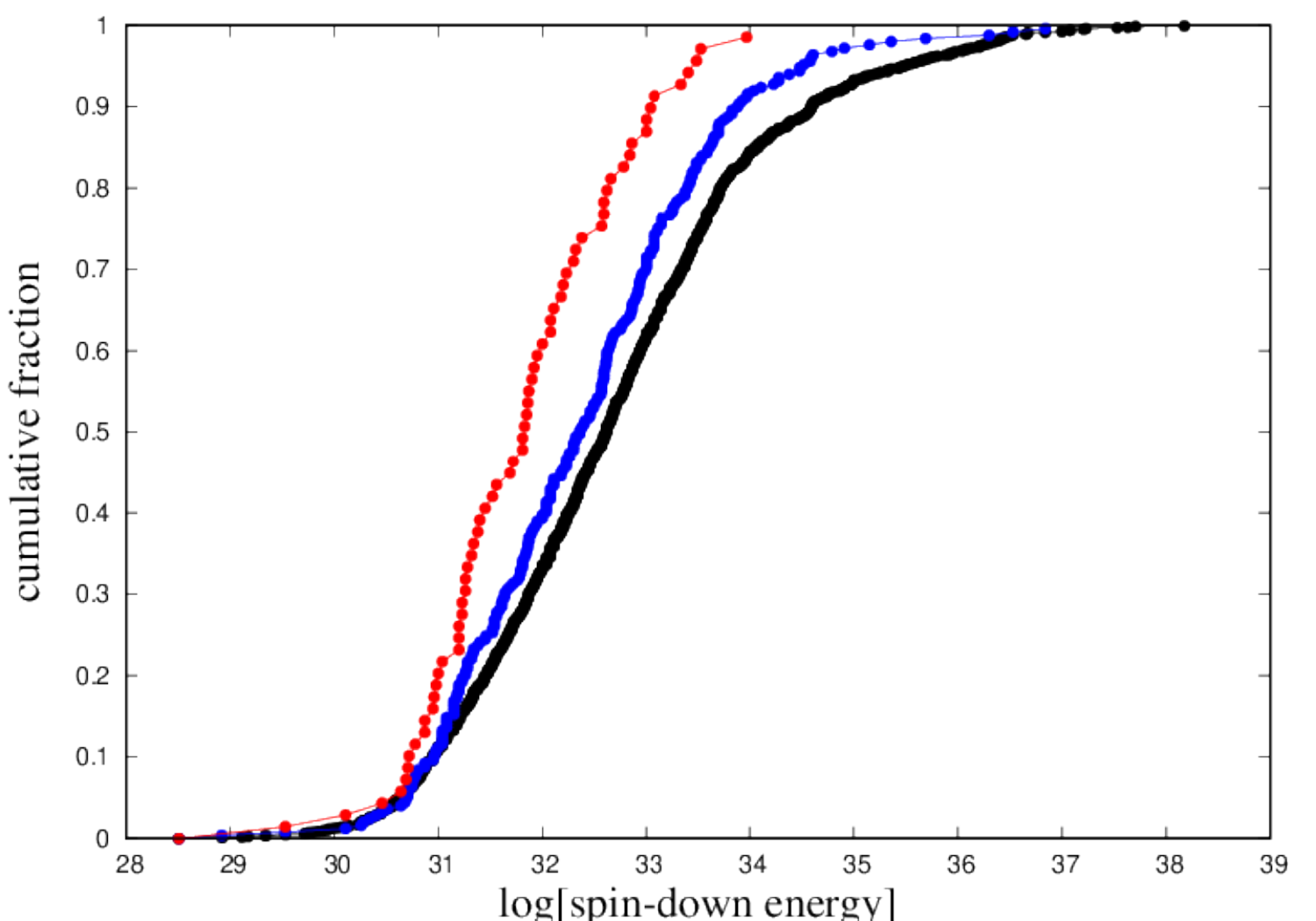}  \\
\end{tabular}
\end{center}
\caption{Cumulative distributions of the log of the spin-down energy, $\dot{E}$, for the 1202 pulsars from the TPA (black), the 249 pulsars used in this sample (blue) and the 69 pulsars with no highly polarized samples (red).}
\label{fig:edot}
\end{figure}

In Figure~\ref{fig:edot} we show the cumulative $\dot{E}$ distributions for the entire TPA sample of 1202 pulsars, the 249 pulsars considered here and the 72 pulsars without any highly polarized samples. Two points can be gleaned. First the 249 pulsars are somewhat lower in $\dot{E}$ than the entire sample; this implies that single pulses from lower $\dot{E}$ pulsars are brighter than those from high $\dot{E}$ pulsars (possibly due to the fact that their distances are smaller). Secondly the pulsars without highly polarized data points are significantly lower in $\dot{E}$ than the sample as a whole. However, we caution that with only $\sim$1000 single pulses per pulsar that some of the seemingly unpolarized cases may be resolved with a much larger number of pulses. This will not be true in all cases, for example the very bright PSR~J1534--5334 shows no samples with linear polarization greater than 50\%.

\subsection{Circular Polarization and OPM jumps}
In the profiles made from the highly linearly polarized samples there often remains a significant degree of circular polarization, for example PSR~J1224--6407 in Figure~\ref{fig1} and PSRs~J0729--1836 and J1312--5516 in Figure~\ref{fig2}. One possibility is that the modes are elliptically polarized. For PSR~J1312--5516 in particular the standard integrated profile with its two competing modes shows almost no circular whereas in the highly polarized profile a single mode dominates with positive circular polarization.

In the highly polarized profiles we find that OPM tracks are present in 35\% of cases, a number not too dissimilar to that found by J23 in a much larger sample of integrated pulse profiles. For the pulsars which do show evidence of OPM jumps in J23, 50\% of them retain the OPM jumps in the highly polarized profiles. It therefore seems clear that sometimes the X mode provides the highly polarized samples, sometimes in the O mode and sometimes a mixture of the two as a function of pulse longitude. In regions of pulse longitude where there are mode mixtures, we find that the average PPA tends to follow one of the modes, and the OPM jumps in the average PPA can be clearly distinguished. The RVM fitter is designed to detect OPM jumps of 90\degr, and hence fitting RVM to the average PPA performs well. For a given pulsar there is no immediate way to tell which of O or X is the dominant mode. Thus, any theoretical emission model should take this into account.

\begin{figure}
\begin{center}
\begin{tabular}{c}
\includegraphics[width=8cm]{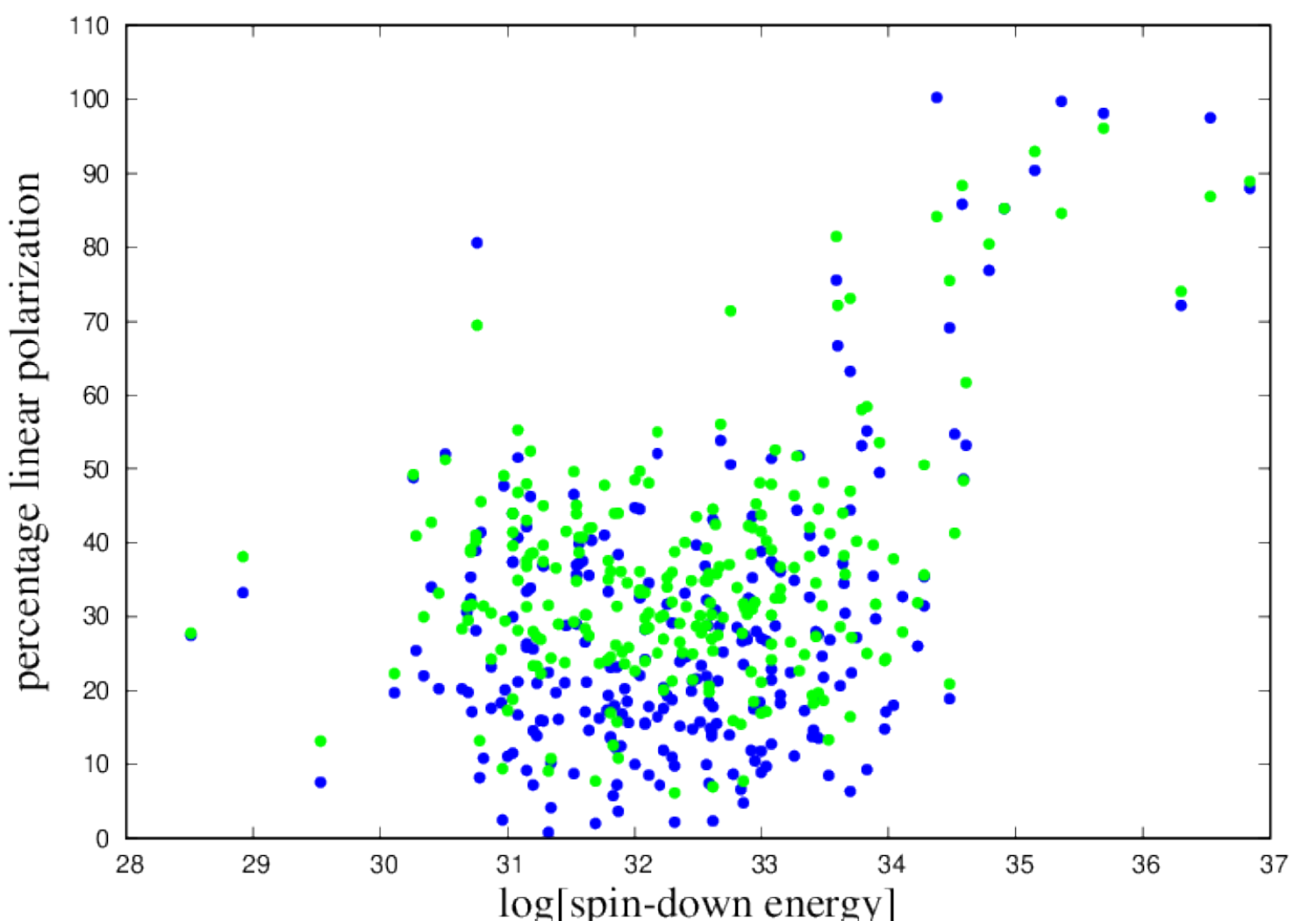}  \\
\end{tabular}
\end{center}
\caption{Percentage linear polarization as a function of $\dot{E}$. The blue points show the linear polarization of the integrated profile, the green points show the mean linear polarization of the individual samples..}
\label{fig:edotvsl}
\end{figure}

\subsection{Tabular description of results}
\label{sectable}
In Table~\ref{tab:all} we present the results for the 249 pulsars analysed here using the selection criteria outlined in Section~\ref{sec2}. The first two columns report the pulsar's Jname and the log of the spin-down energy ($\dot{E}$). Columns 3 and 4 given the number of single pulses recorded ($N_p$) and the percentage of those above 5-$\sigma$ ($N_f$). Columns 5 and 6 give the percentage of linear ($\%L_t$) and circular ($\%V_t$) polarization in the integrated profile. Note that these values may differ slightly from those given in \citet{2023MNRAS.520.4582P} as the observations used here are not necessarily the same as used in their paper. Columns 7 and 8 give the mean percentage of all the samples in linear ($\%L_s$) and circular ($\%V_s$) polarization provided that the total intensity of the sample is greater than 3.5-$\sigma$. The last two columns give the class from the RVM fitting according to J23 and after formation of the high linear profile. We denote with a dagger symbol those pulsars which remain non-RVM even after the application of a higher linear threshold as outlined in Section~\ref{secrvm}.

Figure~\ref{fig:edotvsl} shows the percentage linear polarization of the integrated profile (blue points) and the average of the individual samples (green points). First it should be noted the relative paucity of points above an $\dot{E}$ of $10^{34}$~ergs$^{-1}$. This is because these pulsars are generally too weak to satisfy our single pulse thresholds. However, the blue points show the trend pointed out by \citet{2008MNRAS.391.1210W}, \citet{2023MNRAS.520.4582P} and others in that pulsars with high $\dot{E}$ are highly polarized whereas pulsars at low $\dot{E}$ much less so. In the single samples we see, as expected, that the fractional linear polarization increases. The trend still remains; relatively few pulsars at low $\dot{E}$ have a fractional polarization greater than 50\%. This implies that the majority of samples must have fractional linear polarization less than 25\% and that the modes have already mixed on this short time-scale. 

\section{Conclusions}
\label{sec4}
MMB23 suggest that if the underlying emission mechanism is CCR, then high linearly polarized samples in pulsars should have orthogonal PPA tracks that follow the RVM. We tested this hypothesis on the sample set of TPA pulsars. We found that in pulsars with a disordered PPA distribution, the RVM can be recovered in the vast majority of cases. Further, we clearly see examples of high linearly polarized OPM tracks which can be interpreted as the X and O modes of the strongly magnetized pair plasma. The escape of the X and O modes requires the plasma flow to be inhomogeneous and \cite{2023ApJ...952..151M} alluded to the fact that the observed change in degree of linear polarization with pulsar spindown energy may be related to the change in the degree of inhomogeneity in the plasma.

\cite{2023ApJ...952..151M} suggested that as the linearly polarized modes propagate through the magnetospheric plasma, the modes can get partially converted into circular polarization due to propagation effects, and thus the observed signals are generally elliptically polarized. The non-RVM PPAs can be interpreted as incoherent averaging of the X and O mode from a large number of sources. A different idea was put forward by \citet{2023MNRAS.525..840O}. In their picture, both coherent and incoherent mixing of the modes take place, the coherent mixing produces the circular polarization and also distorts the PPA. This model provides a good fit to the broad-band data presented in \citet{2023MNRAS.520.4961O}.

CCR can be excited in a pulsar plasma by a stable charge bunch streaming at relativistic speeds along the 
curved magnetic field. Theoretically, formation of relativistic charge solitons is thought to be an excellent candidate for the formation of such charge bunches (\citealt{1980Afz....16..161M,1998MNRAS.301...59A,2000ApJ...544.1081M,2018MNRAS.480.4526L,2020MNRAS.497.3953R,2021A&A...649A.145M,2022MNRAS.516.3715R}). The theory of charge solitons has been developed in the one dimensional approximation, and further improvement in the theory in needed to verify the stability of the solitons in a more realistic situation of two or three dimension. Further the physical process of the escape of X and O mode in pulsar plasma and the origin of circular polarization still needs to be understood (e.g. \citealt{2014ApJ...794..105M,2022MNRAS.512.3589R,2023ApJ...952..151M}). The observational evidence for CCR presented in this work strongly motivates theoretical research in these directions. 

\section*{Acknowledgements}
DM acknowledges CSIRO for supporting a visiting position, the Department of Atomic Energy, Government of India, under project No. 12-R\&D-TFR-5.02-0700 and grant 2020/37/B/ST9/02215 from the National Science Centre, Poland. LSO acknowledges the support of Magdalen College, Oxford. We thank M Rahaman for useful comments. 
The MeerKAT telescope is operated by the South African Radio Astronomy Observatory (SARAO), which is a facility of the National Research Foundation, an agency of the Department of Science and Innovation. 
SARAO acknowledges the ongoing advice and calibration of GPS systems by the National Metrology Institute of South Africa (NMISA) and the time space reference systems department department of the Paris Observatory. 
PTUSE was developed with support from the Australian SKA Office and Swinburne University of Technology. 
This work made use of the OzSTAR national HPC facility at Swinburne University of Technology.
MeerTime data is housed on the OzSTAR supercomputer.
The OzSTAR program receives funding in part from the Astronomy National Collaborative Research Infrastructure Strategy (NCRIS) allocation provided by the Australian Government.
\section*{Data Availability}
Data are available on reasonable request. 



\bibliographystyle{mnras}
\bibliography{References} 


\appendix
\section{Table of Results}
Table~\ref{tab:all} lists the 249 pulsars used in this analysis. For a description of the table see Section~\ref{sectable}.

\begin{table*}
\caption{Results for 249 pulsars. The first two columns report the pulsar's Jname and the log of the spin--down energy ($\dot{E}$). Columns 3 and 4 given the number of single pulses recorded ($N_p$) and the percentage of those above 5--$\sigma$ ($N_f$). Columns 5 and 6 give the percentage of linear ($\%L_t$) and circular ($\%V_t$) polarization in the integrated profile. Columns 7 and 8 give the mean percentage of all the samples in linear ($\%L_s$) and circular ($\%V_s$) polarization. The last two columns give the Class from the RVM fitting according to Johnston et al. (2023) and after formation of the high linear profile. The dagger symbol denotes pulsars which remain non--RVM even after application of a higher linear threshold.}
\label{tab:all}
\begin{center}
\begin{tabular}{cccrrrrrrcc}
\hline
\hline
JNAME & log($\dot{E}$) & $N_p$ & $N_f$ & $\%L_t$ & $\%V_t$ & $\%L_s$ & $\%V_s$ & Old Class & New Class\\
\hline
J0034--0721 & 31.3 & 1042 & 54.5 & 15.9 & 4.4 & 37.4 & 4.0 & non--RVM & RVM \\
J0108--1431 & 30.8 & 1042 & 55.6 & 80.6 & 12.1 & 69.5 & 7.4 & flat & flat \\
J0134--2937 & 33.1 & 1110 & 88.1 & 51.4 & --20.3 & 47.9 & --13.7 & RVM & RVM \\
J0151--0635 & 30.7 & 1039 & 97.6 & 38.9 & 0.2 & 41.0 & --1.0 & RVM & RVM \\
J0152--1637 & 31.9 & 1047 & 86.4 & 15.7 & --1.4 & 25.8 & --1.6 & non--RVM & No poln \\
J0255--5304 & 31.1 & 1226 & 81.9 & 9.2 & --5.0 & 31.3 & --4.5 & non--RVM & non--RVM \\
J0302+2252 & 30.3 & 1030 & 57.2 & 25.4 & --0.2 & 40.9 & --0.1 & non--RVM & non--RVM \\
J0304+1932 & 31.3 & 1032 & 89.2 & 36.8 & 12.5 & 45.0 & 12.6 & RVM & RVM \\
J0401--7608 & 32.6 & 1078 & 98.1 & 28.8 & --2.3 & 34.8 & --2.4 & non--RVM & RVM \\
J0448--2749 & 31.8 & 1048 & 58.1 & 23.1 & --15.3 & 36.1 & --17.3 & flat & flat \\
J0452--1759 & 33.1 & 1042 & 100.0 & 18.3 & --0.0 & 33.7 & 0.5 & RVM & RVM \\
J0517+2212 & 31.6 & 1036 & 83.0 & 21.1 & 6.1 & 30.2 & 6.7 & non--RVM & RVM \\
J0525+1115 & 31.8 & 1048 & 95.1 & 13.8 & 9.1 & 24.5 & 9.2 & non--RVM & No poln \\
J0536--7543 & 31.1 & 1041 & 92.6 & 51.5 & --11.6 & 55.3 & --10.9 & RVM & RVM \\
J0543+2329 & 34.6 & 1033 & 97.8 & 53.3 & --11.1 & 61.7 & --12.2 & RVM & RVM \\
J0601--0527 & 32.9 & 1063 & 96.9 & 32.1 & 1.5 & 38.4 & --0.5 & non--RVM & RVM \\
J0614+2229 & 34.8 & 1032 & 100.0 & 76.9 & 16.2 & 80.5 & 15.9 & RVM & RVM \\
J0624--0424 & 31.5 & 1056 & 97.5 & 28.8 & 6.3 & 41.5 & 6.1 & non--RVM & non--RVM \\
J0629+2415 & 32.9 & 1030 & 100.0 & 23.5 & 11.1 & 31.7 & 11.7 & non--RVM & RVM \\
J0630--2834 & 32.2 & 1043 & 100.0 & 52.1 & --5.1 & 55.1 & --4.9 & RVM & RVM \\
J0646+0905 & 31.6 & 1046 & 87.4 & 37.5 & --3.8 & 40.7 & --3.8 & RVM & RVM \\
J0659+1414 & 34.6 & 1047 & 63.4 & 85.8 & --14.0 & 88.4 & --14.4 & RVM & RVM \\
J0719--2545 & 32.5 & 1043 & 48.1 & 21.6 & 8.3 & 28.7 & 8.3 & non--RVM & RVM \\
J0729--1836 & 33.7 & 1042 & 80.7 & 27.2 & --8.3 & 40.1 & --8.2 & RVM & RVM \\
J0738--4042 & 33.0 & 1095 & 100.0 & 27.1 & --6.2 & 43.7 & --6.0 & non--RVM & RVM \\
J0742--2822 & 35.1 & 2969 & 100.0 & 90.4 & --4.8 & 93.0 & --4.9 & RVM & RVM \\
J0758--1528 & 32.3 & 1044 & 79.3 & 18.8 & --1.1 & 32.0 & --0.8 & non--RVM & RVM \\
J0809--4753 & 32.9 & 1044 & 99.8 & 27.4 & --0.9 & 31.2 & --0.8 & non--RVM & RVM \\
J0818--3232 & 30.5 & 1029 & 79.6 & 20.2 & --3.8 & 33.1 & --2.9 & non--RVM & No poln \\
J0820--1350 & 31.6 & 1056 & 98.0 & 14.6 & --9.7 & 27.4 & --10.3 & non--RVM & RVM \\
J0820--4114 & 30.7 & 1048 & 93.9 & 30.6 & 3.3 & 31.4 & 2.2 & No poln & RVM \\
J0823+0159 & 30.8 & 1076 & 99.3 & 10.9 & --2.3 & 31.4 & --1.8 & non--RVM & RVM \\
J0835--4510 & 36.8 & 3433 & 100.0 & 88.0 & --9.0 & 88.9 & --9.3 & RVM & RVM \\
J0837+0610 & 32.1 & 1037 & 92.4 & 8.6 & --4.4 & 28.5 & --4.2 & non--RVM & No poln \\
J0837--4135 & 32.5 & 1041 & 99.5 & 15.8 & 12.4 & 30.1 & 12.5 & non--RVM & RVM \\
J0840--5332 & 32.2 & 1035 & 97.0 & 17.6 & 14.1 & 27.0 & 14.5 & non--RVM & RVM \\
J0846--3533 & 31.7 & 1033 & 99.9 & 40.3 & --10.4 & 42.0 & --10.8 & non--RVM & RVM \\
J0855--3331 & 32.1 & 1030 & 64.2 & 15.5 & 2.2 & 28.3 & 2.4 & non--RVM & No poln \\
J0856--6137 & 31.9 & 1033 & 86.6 & 3.7 & 5.0 & 10.9 & 4.8 & RVM & No poln \\
J0902--6325 & 31.2 & 1033 & 97.6 & 33.9 & 6.9 & 38.5 & 8.9 & non--RVM & RVM \\
J0904--7459 & 32.0 & 1040 & 86.0 & 32.6 & 2.1 & 33.6 & 1.5 & RVM & RVM \\
J0905--4536 & 30.8 & 1030 & 89.3 & 41.4 & 2.5 & 45.5 & 1.4 & RVM & RVM \\
J0905--5127 & 34.4 & 1064 & 2.4 & 100.2 & 8.2 & 84.2 & 9.4 & RVM & RVM \\
J0907--5157 & 33.6 & 1073 & 100.0 & 37.1 & 3.1 & 44.0 & 2.3 & RVM & RVM \\
J0908--1739 & 32.6 & 1050 & 61.9 & 13.8 & 1.9 & 30.2 & 2.9 & non--RVM & RVM \\
J0908--4913 & 35.7 & 1045 & 100.0 & 98.1 & 1.1 & 96.1 & 1.0 & RVM & RVM \\
J0909--7212 & 30.7 & 1035 & 70.8 & 32.4 & 9.2 & 38.7 & 8.0 & RVM & RVM \\
J0922+0638 & 33.8 & 2090 & 100.0 & 55.2 & 8.6 & 58.5 & 8.5 & RVM & RVM \\
J0924--5302 & 33.5 & 1032 & 92.8 & 8.5 & --5.9 & 13.3 & --6.3 & non--RVM & No poln \\
J0924--5814 & 32.7 & 1044 & 99.9 & 53.9 & --3.9 & 56.1 & --4.2 & RVM & RVM \\
J0934--5249 & 31.8 & 1031 & 98.5 & 19.4 & --12.8 & 35.0 & --12.4 & non--RVM & RVM \\
J0942--5552 & 33.5 & 1040 & 98.1 & 38.9 & --2.6 & 48.1 & 0.8 & non--RVM & RVM \\
J0944--1354 & 31.0 & 1041 & 98.7 & 20.1 & 16.0 & 29.4 & 15.4 & non--RVM & No poln \\
J0953+0755 & 32.7 & 803 & 99.8 & 14.0 & --5.6 & 37.0 & --6.5 & non--RVM & RVM \\
\hline
\end{tabular}
\end{center}
\end{table*}
\begin{table*}
\addtocounter{table}{-1}
\caption{Results for 249 pulsars (continued)}
\begin{center}
\begin{tabular}{cccrrrrrrcc}
\hline
\hline
JNAME & log($\dot{E}$) & $N_p$ & $N_f$ & $\%L_t$ & $\%V_t$ & $\%L_s$ & $\%V_s$ & Old Class & New Class\\
\hline
J0959--4809 & 31.0 & 1049 & 95.3 & 43.9 & --7.8 & 39.6 & --10.6 & RVM & RVM \\
J1001--5507 & 32.8 & 1030 & 100.0 & 6.7 & --0.6 & 15.4 & --0.9 & non--RVM & No poln \\
J1001--5559 & 30.9 & 1031 & 88.8 & 23.2 & 10.2 & 30.5 & 13.1 & non--RVM & No poln \\
J1001--5939 & 30.7 & 517 & 43.3 & 35.3 & 1.2 & 39.1 & 0.8 & RVM & No poln \\
J1016--5345 & 32.2 & 1033 & 51.8 & 20.4 & 3.7 & 30.1 & 3.8 & RVM & RVM \\
J1017--5621 & 33.0 & 1046 & 73.3 & 18.4 & --17.1 & 48.1 & --17.4 & non--RVM & non--RVM$\dagger$ \\
J1018--1642 & 31.1 & 1029 & 61.7 & 16.7 & 20.3 & 34.9 & 24.0 & non--RVM & non--RVM \\
J1034--3224 & 30.8 & 1035 & 100.0 & 8.2 & 5.8 & 13.2 & 4.5 & non--RVM & No poln \\
J1041--1942 & 31.1 & 1055 & 98.3 & 33.4 & 7.6 & 43.0 & 8.8 & RVM & RVM \\
J1042--5521 & 32.2 & 1032 & 86.1 & 11.9 & 3.9 & 20.1 & --6.3 & RVM & No poln \\
J1048--5832 & 36.3 & 7270 & 85.3 & 72.2 & 4.5 & 74.0 & 4.8 & RVM & RVM \\
J1049--5833 & 31.2 & 1035 & 48.9 & 25.6 & --7.6 & 38.6 & --7.7 & RVM & No poln \\
J1056--6258 & 33.3 & 1049 & 100.0 & 44.4 & --2.4 & 51.7 & --2.5 & flat & flat \\
J1057--5226 & 34.5 & 1022 & 100.0 & 69.1 & 0.1 & 75.5 & --0.2 & non--RVM & non--RVM$\dagger$ \\
J1059--5742 & 32.0 & 1054 & 96.3 & 10.0 & --0.4 & 22.6 & 0.1 & non--RVM & No poln \\
J1110--5637 & 32.7 & 1041 & 95.8 & 28.7 & --8.0 & 36.7 & --8.9 & RVM & RVM \\
J1112--6613 & 32.9 & 1033 & 99.9 & 17.6 & 7.2 & 18.5 & 8.0 & non--RVM & RVM \\
J1112--6926 & 32.3 & 1036 & 63.0 & 11.0 & 3.1 & 21.3 & 2.9 & non--RVM & No poln \\
J1114--6100 & 33.4 & 1036 & 83.9 & 18.9 & --0.8 & 27.3 & --1.5 & RVM & RVM \\
J1116--4122 & 32.6 & 1037 & 82.7 & 10.0 & --0.8 & 28.8 & --1.4 & RVM & No poln \\
J1121--5444 & 32.9 & 1033 & 99.3 & 26.7 & --7.3 & 27.7 & --7.7 & non--RVM & RVM \\
J1123--4844 & 32.3 & 1057 & 96.7 & 31.7 & 27.0 & 34.0 & 26.3 & non--RVM & RVM \\
J1133--6250 & 31.2 & 1033 & 92.2 & 13.9 & 2.5 & 23.3 & 1.6 & non--RVM & No poln \\
J1136+1551 & 31.9 & 1031 & 85.0 & 18.5 & --8.1 & 34.6 & --8.4 & non--RVM & RVM \\
J1136--5525 & 33.8 & 1047 & 98.6 & 9.3 & --2.2 & 25.0 & --1.9 & non--RVM & RVM \\
J1137--6700 & 31.2 & 1096 & 93.6 & 14.6 & --1.9 & 23.3 & --0.9 & flat & No poln \\
J1141--6545 & 33.4 & 5200 & 85.4 & 13.6 & 5.7 & 19.7 & 4.5 & RVM & RVM \\
J1146--6030 & 33.5 & 1067 & 97.5 & 26.9 & 3.3 & 41.2 & 4.2 & RVM & RVM \\
J1157--6224 & 33.4 & 1052 & 97.6 & 32.6 & 10.8 & 38.1 & 8.9 & non--RVM & non--RVM \\
J1202--5820 & 33.0 & 1031 & 98.7 & 28.0 & 7.8 & 45.2 & 7.0 & non--RVM & RVM \\
J1224--6407 & 34.3 & 1071 & 100.0 & 35.4 & 22.5 & 50.5 & 22.9 & non--RVM & RVM \\
J1232--4742 & 28.9 & 962 & 82.1 & 33.2 & --2.9 & 38.1 & --4.1 & flat & flat \\
J1237--6725 & 31.0 & 1030 & 58.5 & 2.5 & 1.2 & 9.4 & 1.0 & non--RVM & No poln \\
J1239+2453 & 31.1 & 1029 & 94.8 & 42.1 & 0.4 & 47.9 & 0.2 & non--RVM & RVM \\
J1243--6423 & 33.5 & 1058 & 99.6 & 24.6 & --3.5 & 31.4 & --2.8 & non--RVM & non--RVM \\
J1246+2253 & 31.5 & 1032 & 50.9 & 8.8 & --2.1 & 29.3 & --3.4 & non--RVM & No poln \\
J1253--5820 & 33.7 & 1068 & 98.8 & 63.2 & 6.8 & 73.1 & 7.0 & RVM & RVM \\
J1257--1027 & 31.8 & 1072 & 97.2 & 33.4 & --4.8 & 37.5 & --3.5 & non--RVM & non--RVM \\
J1305--6455 & 32.9 & 1036 & 98.8 & 35.2 & 10.4 & 31.0 & 13.1 & non--RVM & non--RVM \\
J1306--6617 & 33.3 & 1037 & 95.6 & 17.3 & 7.5 & 24.9 & 7.6 & non--RVM & No poln \\
J1312--5516 & 32.6 & 709 & 94.6 & 21.9 & 0.1 & 39.2 & --0.5 & RVM & RVM \\
J1320--5359 & 34.2 & 1043 & 86.9 & 26.0 & --8.9 & 31.9 & --7.2 & RVM & RVM \\
J1326--5859 & 33.1 & 1048 & 96.5 & 37.5 & 8.6 & 39.0 & 8.7 & non--RVM & RVM \\
J1326--6408 & 32.4 & 1034 & 85.6 & 24.5 & 2.5 & 25.1 & 1.9 & non--RVM & No poln \\
J1326--6700 & 33.1 & 1037 & 99.8 & 36.8 & --5.8 & 52.6 & --6.1 & RVM & RVM \\
J1327--6222 & 33.7 & 1043 & 100.0 & 6.4 & 5.0 & 16.5 & 4.0 & non--RVM & RVM \\
J1327--6301 & 33.9 & 1075 & 96.4 & 29.7 & 0.2 & 31.7 & 2.0 & RVM & RVM \\
J1328--4357 & 32.9 & 1038 & 89.1 & 32.5 & 6.4 & 42.2 & 6.9 & RVM & RVM \\
J1328--4921 & 30.9 & 527 & 85.2 & 17.6 & 6.7 & 24.2 & 3.1 & non--RVM & No poln \\
J1338--6204 & 32.5 & 1034 & 99.8 & 21.4 & 13.3 & 24.9 & 14.2 & non--RVM & RVM \\
J1355--5153 & 32.6 & 933 & 86.2 & 2.4 & --2.1 & 7.0 & --2.9 & non--RVM & No poln \\
J1401--6357 & 33.0 & 1035 & 97.6 & 26.7 & --0.9 & 40.2 & 0.1 & RVM & RVM \\
J1414--6802 & 30.4 & 538 & 68.8 & 34.0 & 11.3 & 42.7 & 11.7 & RVM & RVM \\
J1420--5416 & 31.0 & 1025 & 52.2 & 11.5 & --13.9 & 18.8 & --14.4 & non--RVM & No poln \\
J1424--5822 & 33.5 & 1028 & 49.1 & 21.8 & 3.2 & 18.7 & 4.1 & flat & No poln \\
\hline
\end{tabular}
\end{center}
\end{table*}
\begin{table*}
\addtocounter{table}{-1}
\caption{Results for 249 pulsars (continued)}
\begin{center}
\begin{tabular}{cccrrrrrrcc}
\hline
\hline
JNAME & log($\dot{E}$) & $N_p$ & $N_f$ & $\%L_t$ & $\%V_t$ & $\%L_s$ & $\%V_s$ & Old Class & New Class\\
\hline
J1428--5530 & 32.6 & 1037 & 72.1 & 30.9 & 0.7 & 42.4 & 0.8 & RVM & RVM \\
J1430--6623 & 32.4 & 1040 & 100.0 & 15.2 & 2.1 & 29.1 & 1.6 & non--RVM & RVM \\
J1440--6344 & 32.7 & 1032 & 85.8 & 21.3 & 3.2 & 27.5 & 5.0 & non--RVM & No poln \\
J1453--6413 & 34.3 & 1087 & 92.7 & 31.4 & 4.4 & 35.6 & 4.9 & non--RVM & RVM \\
J1456--6843 & 32.3 & 1075 & 99.8 & 9.8 & 1.5 & 38.7 & --1.0 & non--RVM & non--RVM \\
J1507--4352 & 33.4 & 1043 & 91.7 & 27.9 & --6.2 & 34.5 & --6.5 & RVM & RVM \\
J1507--6640 & 33.0 & 1042 & 50.8 & 8.9 & 7.4 & 21.1 & 6.4 & non--RVM & No poln \\
J1514--4834 & 32.6 & 1043 & 48.5 & 7.5 & --3.4 & 19.8 & --1.8 & non--RVM & No poln \\
J1522--5829 & 33.1 & 1046 & 98.1 & 28.7 & 1.4 & 32.5 & 1.5 & non--RVM & RVM \\
J1527--3931 & 31.7 & 743 & 50.3 & 16.3 & 12.7 & 23.7 & 11.5 & RVM & No poln \\
J1527--5552 & 32.6 & 1030 & 55.2 & 18.4 & 13.9 & 20.6 & 13.4 & non--RVM & No poln \\
J1534--5334 & 31.3 & 1032 & 100.0 & 4.2 & 0.9 & 10.8 & 0.0 & non--RVM & No poln \\
J1535--4114 & 33.3 & 1043 & 62.1 & 51.8 & --11.2 & 22.6 & --62.6 & RVM & RVM \\
J1536--3602 & 31.1 & 1032 & 61.4 & 25.8 & --7.6 & 37.5 & --9.3 & non--RVM & RVM \\
J1539--5626 & 34.1 & 1078 & 89.4 & 32.7 & 1.6 & 27.9 & --0.9 & non--RVM & RVM \\
J1539--6322 & 30.3 & 1104 & 91.1 & 48.8 & 5.4 & 49.2 & 5.9 & RVM & RVM \\
J1543+0929 & 31.6 & 1028 & 98.9 & 26.5 & --4.9 & 30.2 & --9.4 & non--RVM & RVM \\
J1544--5308 & 32.6 & 1087 & 99.8 & 17.8 & --4.1 & 25.4 & --5.5 & non--RVM & RVM \\
J1555--2341 & 32.3 & 1046 & 83.9 & 19.3 & 2.0 & 35.2 & 2.3 & RVM & RVM \\
J1555--3134 & 31.3 & 1045 & 99.8 & 16.0 & 2.9 & 26.9 & 2.7 & non--RVM & No poln \\
J1557--4258 & 32.6 & 1038 & 85.7 & 32.2 & --13.8 & 39.1 & --15.3 & non--RVM & non--RVM \\
J1559--4438 & 33.4 & 1066 & 100.0 & 41.0 & --10.3 & 42.0 & --10.0 & RVM & RVM \\
J1602--5100 & 33.6 & 1030 & 94.3 & 20.6 & 6.0 & 28.6 & --4.4 & non--RVM & RVM \\
J1603--2531 & 33.4 & 2131 & 94.3 & 27.7 & 0.6 & 44.5 & --0.4 & flat & flat \\
J1603--2712 & 32.4 & 1030 & 64.8 & 33.1 & 1.2 & 40.0 & 1.7 & non--RVM & non--RVM \\
J1604--4909 & 33.0 & 1043 & 99.9 & 9.8 & 0.8 & 17.1 & 1.1 & non--RVM & No poln \\
J1604--7203 & 31.9 & 1764 & 93.0 & 20.2 & 5.5 & 23.6 & 4.9 & non--RVM & No poln \\
J1605--5257 & 31.5 & 1037 & 99.8 & 37.0 & 5.1 & 45.0 & 4.9 & RVM & RVM \\
J1607--0032 & 32.2 & 1031 & 95.3 & 7.2 & 1.6 & 29.9 & --0.2 & non--RVM & No poln \\
J1622--4332 & 31.9 & 1039 & 58.5 & 16.8 & 12.4 & 25.2 & 8.8 & non--RVM & No poln \\
J1623--0908 & 31.7 & 1030 & 52.3 & 2.0 & 2.8 & 7.8 & 1.9 & non--RVM & No poln \\
J1625--4048 & 30.1 & 523 & 44.0 & 19.7 & 1.6 & 22.3 & 2.0 & non--RVM & No poln \\
J1635--5954 & 32.6 & 1030 & 97.9 & 36.8 & 8.1 & 34.9 & 8.8 & non--RVM & RVM \\
J1645--0317 & 33.1 & 1062 & 100.0 & 12.8 & --1.2 & 24.2 & --0.7 & non--RVM & RVM \\
J1646--6831 & 31.1 & 1031 & 70.2 & 40.6 & 6.8 & 46.8 & 6.2 & non--RVM & non--RVM$\dagger$ \\
J1650--1654 & 31.4 & 1029 & 66.0 & 19.7 & 7.2 & 36.6 & 4.8 & non--RVM & No poln \\
J1651--4246 & 32.5 & 2143 & 96.1 & 39.6 & --15.4 & 43.5 & --13.7 & non--RVM & RVM \\
J1651--5222 & 32.4 & 1054 & 99.2 & 19.9 & --5.6 & 34.9 & --5.3 & non--RVM & RVM \\
J1700--3312 & 31.9 & 1037 & 76.0 & 38.4 & --15.2 & 44.0 & --14.8 & RVM & RVM \\
J1703--1846 & 32.1 & 1057 & 67.8 & 34.5 & 3.9 & 48.0 & 4.2 & RVM & RVM \\
J1703--3241 & 31.2 & 1040 & 99.7 & 46.2 & --0.5 & 52.5 & --0.6 & RVM & RVM \\
J1705--1906 & 33.8 & 868 & 100.0 & 53.2 & --17.5 & 58.1 & --17.6 & non--RVM & non--RVM$\dagger$ \\
J1705--3423 & 33.4 & 1098 & 97.4 & 13.7 & 1.9 & 19.4 & 2.3 & non--RVM & RVM \\
J1709--1640 & 32.9 & 1033 & 96.0 & 10.5 & 2.0 & 31.9 & 2.4 & RVM & RVM \\
J1709--4429 & 36.5 & 2204 & 93.0 & 97.5 & --21.3 & 86.9 & --21.3 & flat & flat \\
J1711--5350 & 32.9 & 669 & 60.5 & 11.9 & --2.3 & 22.5 & --0.5 & non--RVM & RVM \\
J1720--1633 & 31.8 & 1033 & 76.8 & 17.3 & 3.5 & 24.1 & 2.1 & RVM & RVM \\
J1720--2933 & 32.1 & 1048 & 69.8 & 15.6 & 5.7 & 29.8 & 5.2 & RVM & No poln \\
J1722--3207 & 32.4 & 1068 & 100.0 & 23.9 & 1.1 & 26.6 & 0.7 & RVM & RVM \\
J1722--3712 & 34.5 & 1100 & 34.5 & 54.8 & 12.4 & 41.3 & 12.2 & RVM & RVM \\
J1731--4744 & 34.0 & 1049 & 100.0 & 18.0 & 4.0 & 37.8 & 3.7 & non--RVM & non--RVM$\dagger$ \\
J1735--0724 & 32.8 & 1072 & 91.0 & 28.5 & 0.9 & 33.9 & 1.9 & non--RVM & RVM \\
J1738--3211 & 31.8 & 1054 & 52.4 & 17.9 & --5.0 & 44.0 & --5.5 & non--RVM & non--RVM$\dagger$ \\
J1740+1000 & 35.4 & 1311 & 77.0 & 99.7 & --3.1 & 84.6 & --2.2 & RVM & RVM \\
J1740+1311 & 32.0 & 1039 & 98.9 & 44.5 & 4.1 & 49.7 & 4.0 & RVM & RVM \\
\hline
\end{tabular}
\end{center}
\end{table*}
\begin{table*}
\addtocounter{table}{-1}
\caption{Results for 249 pulsars (continued)}
\begin{center}
\begin{tabular}{cccrrrrrrcc}
\hline
\hline
JNAME & log($\dot{E}$) & $N_p$ & $N_f$ & $\%L_t$ & $\%V_t$ & $\%L_s$ & $\%V_s$ & Old Class & New Class\\
\hline
J1740--3015 & 34.9 & 1032 & 99.7 & 85.3 & --34.0 & 85.3 & --34.9 & RVM & RVM \\
J1741--0840 & 31.0 & 1037 & 71.8 & 37.4 & --4.4 & 44.0 & --4.1 & RVM & RVM \\
J1741--3927 & 32.7 & 1055 & 100.0 & 25.2 & 4.5 & 29.9 & 4.7 & non--RVM & RVM \\
J1742--4616 & 31.3 & 1035 & 38.5 & 36.9 & --8.9 & 39.6 & --8.9 & flat & No poln \\
J1743--3150 & 32.5 & 1035 & 89.1 & 23.4 & --3.7 & 27.8 & --3.0 & RVM & RVM \\
J1744--1610 & 31.2 & 569 & 62.4 & 21.0 & --2.6 & 27.2 & --2.4 & non--RVM & No poln \\
J1745--3040 & 33.9 & 1060 & 78.5 & 49.5 & --2.7 & 53.6 & --3.5 & non--RVM & non--RVM \\
J1748--1300 & 32.9 & 1041 & 93.8 & 26.9 & 3.5 & 30.3 & 4.5 & RVM & RVM \\
J1750--3157 & 31.0 & 1043 & 59.4 & 11.1 & --2.7 & 17.3 & --3.3 & non--RVM & No poln \\
J1751--4657 & 32.1 & 1050 & 97.0 & 17.9 & 9.1 & 30.5 & 8.7 & non--RVM & RVM \\
J1752--2806 & 33.3 & 1050 & 99.9 & 11.1 & 2.8 & 36.6 & 4.6 & non--RVM & non--RVM$\dagger$ \\
J1759--3107 & 32.1 & 1032 & 45.8 & 24.2 & 1.0 & 33.2 & --3.6 & RVM & RVM \\
J1801--2920 & 32.0 & 1029 & 78.3 & 44.7 & --6.9 & 48.5 & --6.6 & RVM & RVM \\
J1806--1154 & 32.6 & 1030 & 94.3 & 35.6 & --0.3 & 35.8 & --0.8 & RVM & RVM \\
J1807--0847 & 32.4 & 1123 & 99.9 & 24.6 & 5.6 & 31.3 & 6.0 & non--RVM & RVM \\
J1808--0813 & 31.9 & 1034 & 80.2 & 23.1 & 6.8 & 31.4 & 7.0 & RVM & RVM \\
J1810--5338 & 32.9 & 2300 & 99.9 & 43.5 & --0.3 & 42.0 & --1.4 & RVM & RVM \\
J1817--3618 & 33.1 & 1063 & 68.8 & 19.4 & --12.4 & 32.5 & --13.2 & RVM & RVM \\
J1817--3837 & 32.6 & 1068 & 96.8 & 14.9 & 0.7 & 31.8 & 1.1 & non--RVM & RVM \\
J1819+1305 & 31.1 & 1034 & 60.2 & 21.2 & --1.9 & 28.1 & --2.8 & RVM & RVM \\
J1820--0427 & 33.1 & 1138 & 94.1 & 22.9 & --6.2 & 30.2 & --6.1 & non--RVM & RVM \\
J1823+0550 & 31.3 & 1032 & 100.0 & 22.4 & 3.9 & 31.5 & 3.8 & non--RVM & RVM \\
J1823--3106 & 33.7 & 1201 & 91.6 & 44.4 & --4.9 & 47.0 & --4.6 & RVM & RVM \\
J1824--1945 & 34.5 & 1112 & 97.7 & 18.9 & --2.3 & 20.9 & --2.6 & non--RVM & RVM \\
J1825--0935 & 33.7 & 1031 & 99.9 & 34.5 & 4.4 & 38.2 & 3.7 & non--RVM & non--RVM$\dagger$ \\
J1826--1131 & 31.3 & 1032 & 87.0 & 0.8 & --6.3 & 9.1 & --4.2 & non--RVM & No poln \\
J1829--1751 & 33.9 & 685 & 100.0 & 35.5 & --9.3 & 39.7 & --9.6 & non--RVM & flat \\
J1832--0827 & 34.0 & 1052 & 81.3 & 14.8 & --2.3 & 24.1 & --0.9 & non--RVM & No poln \\
J1834--0426 & 32.1 & 1071 & 100.0 & 28.2 & --1.8 & 23.9 & --2.3 & No poln & RVM \\
J1836--1008 & 33.4 & 1041 & 100.0 & 14.7 & 5.2 & 18.3 & 5.2 & non--RVM & No poln \\
J1837--0653 & 30.6 & 1029 & 54.4 & 20.2 & --4.2 & 28.3 & --7.5 & non--RVM & No poln \\
J1839--1238 & 31.4 & 1034 & 66.7 & 21.1 & 7.2 & 23.8 & 6.9 & RVM & No poln \\
J1840--0809 & 32.0 & 1030 & 86.3 & 32.6 & --10.7 & 36.1 & --9.7 & RVM & RVM \\
J1840--0815 & 31.9 & 1029 & 78.0 & 7.3 & --0.9 & 15.8 & --0.1 & non--RVM & No poln \\
J1842--0359 & 30.5 & 1035 & 98.9 & 52.1 & 3.7 & 51.2 & 3.3 & RVM & RVM \\
J1843--0000 & 32.7 & 1028 & 100.0 & 15.5 & --0.6 & 35.8 & --0.8 & RVM & RVM \\
J1844+1454 & 33.1 & 1030 & 96.0 & 36.1 & 0.5 & 36.7 & 0.5 & non--RVM & RVM \\
J1847--0402 & 34.0 & 1054 & 99.7 & 17.1 & 2.2 & 24.3 & 2.3 & non--RVM & RVM \\
J1848--0123 & 32.9 & 1047 & 100.0 & 4.8 & 0.9 & 7.8 & 1.0 & non--RVM & No poln \\
J1848--1150 & 31.4 & 518 & 39.8 & 16.1 & 7.9 & 29.0 & 7.2 & non--RVM & No poln \\
J1849--0636 & 32.8 & 1032 & 52.3 & 8.7 & 2.6 & 15.9 & 2.7 & non--RVM & No poln \\
J1852--0635 & 33.6 & 1048 & 96.8 & 66.7 & --1.4 & 72.2 & --1.4 & RVM & RVM \\
J1854--1421 & 32.0 & 575 & 88.5 & 22.0 & 0.1 & 33.2 & 0.9 & RVM & RVM \\
J1900--2600 & 31.5 & 1062 & 95.7 & 35.6 & --0.1 & 43.9 & 0.1 & non--RVM & RVM \\
J1900--7951 & 31.5 & 515 & 67.6 & 29.0 & --1.6 & 34.8 & --2.6 & non--RVM & RVM \\
J1901+0331 & 33.0 & 1034 & 100.0 & 38.8 & 4.6 & 41.5 & 4.8 & non--RVM & RVM \\
J1901--0906 & 31.0 & 1034 & 96.0 & 30.0 & 0.3 & 41.4 & --1.7 & non--RVM & RVM \\
J1902+0556 & 33.1 & 1031 & 80.1 & 21.4 & --4.9 & 26.3 & --3.5 & non--RVM & No poln \\
J1902+0615 & 33.0 & 1035 & 63.9 & 11.8 & 4.5 & 16.9 & 4.0 & RVM & No poln \\
J1903+0135 & 32.6 & 1033 & 99.7 & 14.5 & 9.3 & 27.0 & 9.3 & non--RVM & RVM \\
J1903--0632 & 33.2 & 1388 & 48.7 & 22.4 & 3.8 & 26.6 & 4.7 & RVM & RVM \\
J1909+0007 & 32.3 & 1277 & 71.6 & 2.2 & 2.4 & 6.2 & 3.9 & non--RVM & No poln \\
J1909+1102 & 33.7 & 1053 & 62.1 & 30.4 & --14.7 & 35.7 & --18.5 & non--RVM & RVM \\
J1910+0358 & 31.1 & 1033 & 94.5 & 26.3 & --11.6 & 36.8 & --11.4 & non--RVM & RVM \\
J1912+2104 & 31.6 & 1029 & 74.6 & 39.8 & --11.3 & 38.7 & --8.5 & RVM & RVM \\
\hline
\end{tabular}
\end{center}
\end{table*}
\begin{table*}
\addtocounter{table}{-1}
\caption{Results for 249 pulsars (continued)}
\begin{center}
\begin{tabular}{cccrrrrrrcc}
\hline
\hline
JNAME & log($\dot{E}$) & $N_p$ & $N_f$ & $\%L_t$ & $\%V_t$ & $\%L_s$ & $\%V_s$ & Old Class & New Class\\
\hline
J1913--0440 & 32.5 & 1053 & 100.0 & 14.8 & --1.1 & 21.4 & --1.1 & non--RVM & RVM \\
J1914+0219 & 32.6 & 1046 & 80.0 & 43.1 & --3.1 & 44.5 & 0.1 & RVM & RVM \\
J1915+0738 & 31.6 & 1033 & 39.8 & 37.1 & --18.7 & 40.7 & --17.5 & RVM & No poln \\
J1917+1353 & 34.6 & 1058 & 99.9 & 48.5 & --8.6 & 48.4 & --8.3 & RVM & RVM \\
J1919+0021 & 32.2 & 1037 & 73.3 & 16.4 & 5.5 & 25.1 & 5.9 & non--RVM & No poln \\
J1919+0134 & 30.7 & 1033 & 92.9 & 28.1 & --9.4 & 40.2 & --9.6 & non--RVM & RVM \\
J1921+1948 & 31.8 & 1034 & 85.5 & 13.6 & --2.7 & 17.0 & --6.7 & non--RVM & No poln \\
J1921+2153 & 31.3 & 1030 & 100.0 & 10.2 & 1.6 & 24.4 & 1.7 & non--RVM & RVM \\
J1926+0431 & 31.9 & 558 & 69.2 & 12.5 & 1.1 & 36.1 & 0.9 & non--RVM & RVM \\
J1932+1059 & 33.6 & 1033 & 100.0 & 75.6 & --2.7 & 81.5 & --3.3 & RVM & RVM \\
J1935+1616 & 33.7 & 1033 & 100.0 & 22.4 & --0.8 & 27.2 & --0.0 & non--RVM & RVM \\
J1941--2602 & 32.8 & 1067 & 78.2 & 50.6 & --12.4 & 71.4 & --13.3 & non--RVM & RVM \\
J1943--1237 & 31.9 & 1046 & 63.5 & 12.3 & --2.4 & 26.1 & --2.1 & RVM & No poln \\
J1945--0040 & 31.3 & 1147 & 86.1 & 15.9 & 7.0 & 22.3 & 7.9 & non--RVM & No poln \\
J1946+1805 & 31.0 & 1042 & 40.1 & 37.4 & --6.6 & 43.9 & --6.9 & flat & flat \\
J1946--2913 & 31.8 & 1035 & 73.7 & 5.8 & 3.5 & 12.6 & 2.8 & non--RVM & No poln \\
J2006--0807 & 31.0 & 1051 & 53.3 & 47.6 & 2.9 & 49.0 & 2.8 & non--RVM & non--RVM$\dagger$ \\
J2037+1942 & 31.0 & 520 & 58.6 & 18.3 & --1.3 & 25.5 & --1.6 & non--RVM & No poln \\
J2046+1540 & 30.7 & 1035 & 63.8 & 19.8 & 1.9 & 29.5 & --4.7 & non--RVM & No poln \\
J2046--0421 & 31.2 & 1036 & 99.7 & 7.2 & --12.8 & 28.0 & --11.0 & non--RVM & No poln \\
J2048--1616 & 31.8 & 1036 & 89.5 & 41.0 & 3.9 & 47.8 & 4.2 & RVM & RVM \\
J2053--7200 & 32.3 & 1040 & 98.8 & 29.1 & --6.1 & 36.0 & --3.7 & non--RVM & RVM \\
J2136--1606 & 29.5 & 1044 & 45.7 & 7.6 & 8.9 & 13.2 & 10.8 & non--RVM & No poln \\
J2139+2242 & 31.6 & 1030 & 99.5 & 35.6 & --0.9 & 41.9 & --0.3 & RVM & RVM \\
J2144--3933 & 28.5 & 519 & 99.0 & 27.4 & 4.0 & 27.8 & 2.7 & non--RVM & No poln \\
J2215+1538 & 33.3 & 1077 & 99.6 & 34.9 & --0.5 & 46.3 & --0.2 & RVM & RVM \\
J2253+1516 & 30.7 & 1063 & 42.6 & 17.1 & --4.7 & 31.6 & --3.5 & non--RVM & No poln \\
J2307+2225 & 30.3 & 1029 & 83.4 & 22.0 & 3.5 & 29.9 & 1.6 & flat & flat \\
J2330--2005 & 31.6 & 1036 & 83.1 & 17.1 & --8.3 & 28.4 & --7.7 & non--RVM & non--RVM \\
J2346--0609 & 31.5 & 1033 & 70.6 & 46.5 & 11.4 & 49.6 & 10.5 & RVM & RVM \\
 \\
\hline
\end{tabular}
\end{center}
\end{table*}


\bsp	
\label{lastpage}
\end{document}